# Analysis of Bragg Curve Parameters and Lateral Straggle for Proton and Carbon Beams

Fatih EKİNCİ[1], Erkan BOSTANCI[2], Özlem DAĞLI[3] and Mehmet Serdar GÜZEL[2]
fatih.ekinci2@gazi.edu.tr, ebostanci@ankara.edu.tr, ozlemdagli@gazi.edu.tr, mguzel@ankara.edu.tr
1 Physics Department, Gazi University
2 Computer Engineering Department, Ankara University,
3 Gamma Knife Unit, Gazi University

Abstract. Heavy ions have varying effects on the target. The most important factor in comparing this effect is Linear Energy Transfer (LET). Protons and carbons are heavy ions with high LET. Since these ions lose energy through collisions as they move through the tissue, their range is not long. This loss of energy increases along the way, and the maximum energy loss is reached at the end of the range. This whole process is represented by the Bragg curve. The input dose of the Bragg curve, full width at half maximum (FWHM) value, Bragg peak amplitude and position, and Penumbra thickness are important factors in determining which particle is advantageous in tumor treatment. While heavy ions move through the tissue, small deviations occur in their direction of travel due to Coulomb collisions. These small deviations cause lateral straggle in the dose profile. Lateral straggle is important in determining the type and energy of the particle used in tumor treatments close to critical organs. In our study, when the water phantom of protons and carbon beams with different energies is taken into consideration, the input dose, FWHM value, peak amplitude and position, penumbra thickness and lateral straggle are calculated using the TRIM code and the results are compared with Monte Carlo (MC) simulation. It was found that the proton has an average of 63% more FWHM and 53% more Penumbra than the carbon ion. The carbon ion has an average of 28-45 times greater Bragg peak amplitude at the same Bragg peak location than the proton. It was observed that the proton scattered approximately 70% more in lateral straggle. The difference was found to be around 1.32 mm. In line with all these results, the most commonly used proton and carbon heavy ions in hadron therapy applications were compared.

1. Introduction

Recently, interest in radiation therapy with heavy ions such as protons and carbon has gained momentum. In 2019, more than 200,000 patients were treated with protons in 110 centers around the world and more than 30,000 patients in 13 centers were treated with carbon. It is planned to establish 14 protons and 1 carbon therapy center. It was observed that the greatest proportion of the number of patients and centers treated with Hadron therapy was proton-based. Heavy ion therapy has spread over a wide area, including research institutes or hospitals, since 1980s [1]. The reason why proton and carbon therapy is preferred clinically is that higher doses can be given to the tumor compared to photon radiotherapy and better protection of healthy tissue [2]. The doses given in photon radiotherapy are generally at limited energies to prevent harmful effects on healthy tissue. Whereas, higher doses provide higher tumor control [3]. The majority of the dose is transferred to the tumor with heavy ions such as protons and carbon than photons. Due to the high compatibility and calibration achieved with Hadron therapy, better results have been obtained in treating tumors very close to critical tissues [4].

As the heavy ions from the accelerator move through the tissue, they slow down until they stop and gradually lose their energy. When a charged particle enters the environment, it transfers its energy approximately inversely proportional to the square of its velocity. Therefore, as the particle slows down, the possibility of ionization of the atoms in the environment increases and the maximum LET is transferred to the depth where the ionization events are maximum. All this process loss is represented by the Bragg curve [5-6]. The Bragg curve, calculated using the Bethe-Bloch equation, shows that this decrease in the energy of the particle increases along the way and reaches the maximum energy loss at the end of the range. It appears that the absorbed dose decreases sharply after the peak due to the very small number of particles reaching the back of this peak. Bragg curve; It consists of the Bragg peak, plateau, FWHM (Full Width at Half Maximum), entrance zone and Penumbra [7-8].

Heavily charged particles do not travel in a straight line through the target. There are deviations in their direction due to ionization and collisions in atomic scale. Lateral straggle is a measure of the amount of scatter from the direction of each ion within the target. Lateral straggle occurs mostly at the Bragg peak [8-9]. This concept determines which particle should be used in the treatment of tumors close to critical tissues in hadron therapy.

To the best of the authors' knowledge, it was observed that there is a gap in the comparison of the lateral straggle profiles of protons and carbon ions used in heavy ion therapy. The goal of our study is to find out which of these two particles works better. In this sense, Bragg curve parameters were compared as well as the values of lateral straggle. Thus, effort was made in order to determine which particle will be preferred in tumor treatment close to critical points where lateral straggle is very important.

In this study, the Bragg curves of protons with 80, 100, 120 and 140 MeV energies and carbon bundles with 1.6, 2.4 and 3.0 GeV energies in water were obtained using the TRIM Monte Carlo (MC) simulation software and compared with the literature. After the results, were observed to be compatible with the literature, the Bragg curve parameters and lateral straggle of proton and carbon beams in the water phantom were calculated and compared with each other.

The rest of the paper is structured as follows: Section 2 describes the approach used in the study, followed by Section 3 where the findings are analyzed. A thorough discussion is presented in Section 4 and finally the paper is concluded in Section 5.

2. Methods

MC method is a statistical simulation technique developed for solving mathematical problems where finding an analytical solution is hard. Simulation systems developed on this technique follow the traces of each particle traveling through matter step by step, based on the assumption that the quantities describing particle interactions have certain probability distributions. Quantities such as flux, energy loss and absorbed dose are recorded for many particles and average values for these distributions are computed [10]. TRIM (TRansport of Ions in Matter) simulation software developed using MC technique has the ability to calculate all interactions of ions within the target. The type, energy, target phantom type and shape, parameter to be calculated, particle and probability number of ions can be selected from the TRIM screen. The program records all calculation fields and can view as required [8].

As with photon radiotherapy, the most important problem for hadron therapy is whether the desired dose can be administered to the patient. For this, before the patient is treated, an attempt is made to determine and calibrate the correct dose using the water phantom [11]. Water is the most important medium used in medical physics. Reliability of stopping power calculations for water and accurate calculation of dose distribution mean reliable treatment doses for patients. This is due to the fact that the main component of the human body is considered water. In hadron therapy applications, as in photon radiotherapy, dose distribution is controlled by tissue equivalent phantoms (such as water phantoms). In this respect, the shape and design of the phantom structure to be used are important. There are phantom types used for different body planning in literature [12]. In this study, a cylindrical water phantom was employed.

## 3. Findings

In order to test the accuracy of the calculations in order to find the appropriate doses of protons and carbon beams in the water phantom, the Bragg curves of 80,100, 120 and 140 MeV energy proton beams and 1.6, 2.4 and 3.0 GeV energy carbon beams normalized to the maximum dose in the water phantom were compared with the literature [ 13-23]. By comparison, an average difference of 3.37% for the two particles was observed is generally not significant and is within acceptable limits (<5%) in medical physics. The deviations above the acceptable difference are within acceptable limits in the literature considering the inhomogeneity effects and Monte Carlo-based probabilities.

The energies of the protons and carbon beams were chosen at energies that would have the same Bragg peak positions. According to the Bethe–Bloch equation, the penetration depth (R) of particles with the same kinetic energy is the ratio of the mass number (A) to square of the atomic number (Z); namely $R \sim A/Z^2$. Therefore, one can expect different range values for Protons (A = Z = 1) and carbon particles (A = 12, Z = 6) [24]. Looking at the Bragg curves of these particles in the water phantom (Table 1 and Figure 1), carbon bundles require 12 times more energy for achieving the same range. The input LET was realized as an average of 0.0716 eV / A in the proton beam and an average of 1.6260 eV/A in the carbon beam. As the energy increased, the input LET decreased within two particles. The Bragg peak amplitude (Table 1) was found to be an average of 12.4772 eV/A in the carbon beam and 0.3563 eV/A in the proton. The carbon particle Bragg peak transferred an average of 35 times more LET. The average FWHM value (Figures 1 and 2) was found to be 1.5 cm in the proton and 0.48 cm in the carbon. In penumbra value, the proton was found to be about 0.8 cm, while it was found to be about 0.32 cm in carbon.

**Table 1.** Bragg peak positions, peak amplitudes and percentage differences of protons and carbon beams

| Energy (MeV) | Proton | | Energy (MeV/u) | Carbon | |
|---|---|---|---|---|---|
| | Bragg peak (cm) | Peak Amplitude (eV/A) | | Bragg peak (cm) | Peak Amplitude (eV/A) |
| 80 | 5.2 | 0.4130 | 150 | 5.3 | 11.62910 |
| 100 | 7.6 | 0.3877 | 183 | 7.5 | 12.67340 |
| 120 | 10.4 | 0.3291 | 217 | 10.1 | 13.41200 |
| 140 | 13.6 | 0.2953 | 258 | 13.6 | 12.19430 |

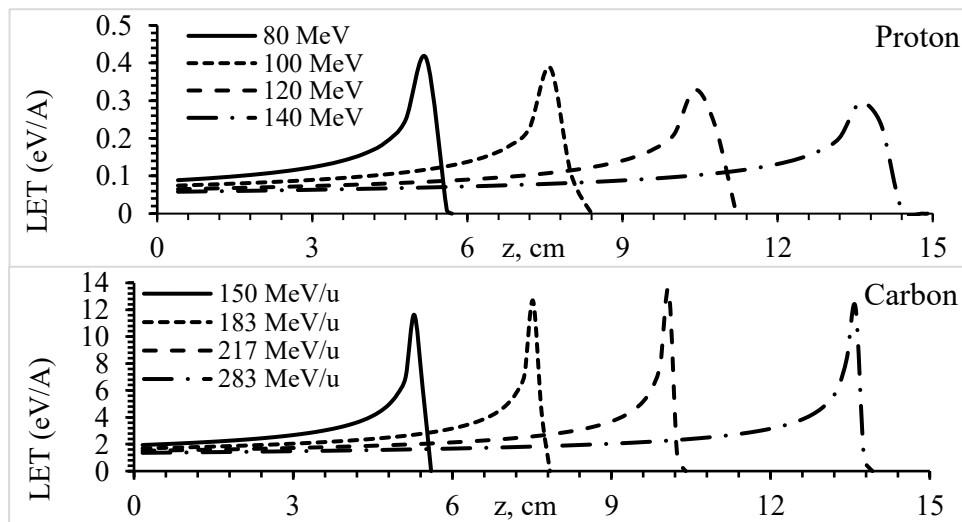

**Figure 1.** Bragg curves for proton and carbon beams

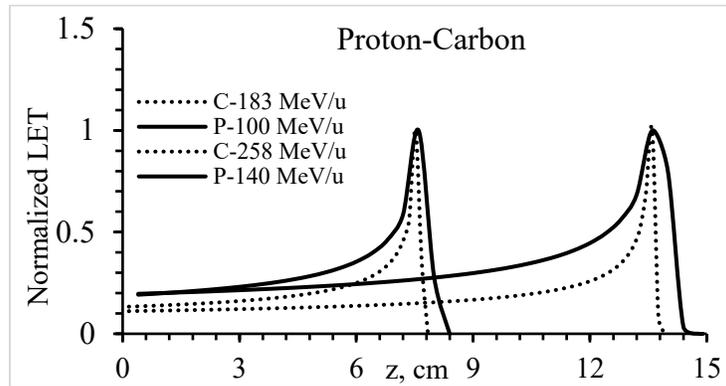

**Figure 2.** FWHM for proton and carbon beams

Lateral straggle (Table 2) was realized as an average of 1.878 mm in the proton and 0.558 mm in carbon. As the energy increased, the lateral straggle increased by 0.470 mm, *i.e.* 59%, in the carbon beam, and 1.77 mm, *i.e.* 64%, in the proton beam. The increase in energy caused an increase in range and therefore an increase in lateral straggle. Lateral straggle occurred more sharply (y = 0.0297x-1.384) for the proton beam (Figure 3).

**Table 2.** Lateral straggle difference for proton and carbon beams

| Energy (GeV/u) | Proton (mm) | Energy (GeV/u) | Carbon (mm) | Difference, mm | Difference, % |
|---|---|---|---|---|---|
| 80 | 1.01 | 150 | 0.33 | 0.68 | 67 |
| 100 | 1.55 | 183 | 0.49 | 1.06 | 68 |
| 120 | 2.17 | 217 | 0.61 | 1.56 | 72 |
| 140 | 2.78 | 250 | 0.80 | 1.98 | 71 |
| | | | **Mean difference** | **1.32** | **70** |

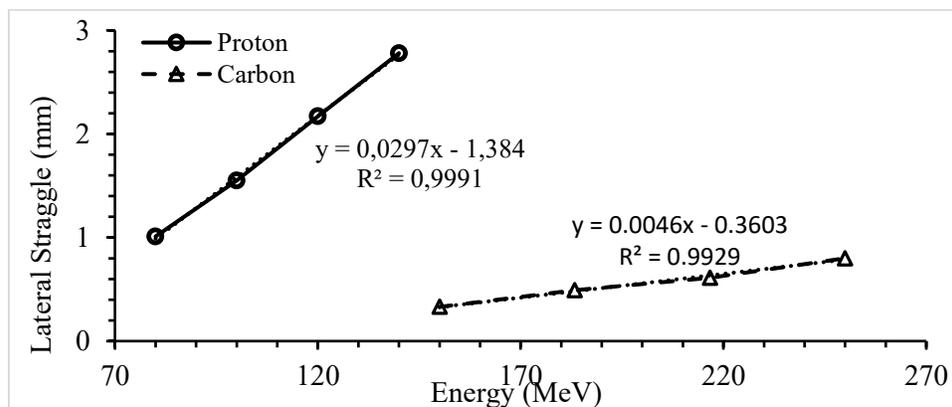

**Figure 3.** Change in the lateral straggle vs energy in water phantom

## 4. Discussion

The physical and radiobiological properties of heavy ions provide a superior dose distribution compared to photon radiotherapy, thus minimizing the dose delivered to normal tissues. Thus, the risk of secondary cancer is significantly reduced [25]. In photon radiotherapy, there are risks from side effects due to the high input dose and a non-zero output dose. In contrast, proton therapy has a significantly lower input dose and little output dose, which reduces damage to healthy tissue surrounding the tumor [26]. Since the use of Hadron therapy, the number of treatable diseases has increased [27].

Hadron therapy has provided a great advantage over photon-based therapy in terms of protecting multiple critical organs [28–29]. In this study, we investigated which of the proton and carbon particle is appropriate to calculate the appropriate dose in the treatment with a comparison based on the Bragg peak properties, especially the lateral straggle results. Studies have found a reduction in toxicity potential after heavy ion therapy [30-31]. In some studies, a significant reduction in the risk of side effects was observed in approximately 70% of cases, relative to the current and confirmed likelihood of normal tissue complications [32]. Especially in head and neck cancer, heavy ion therapy is uniquely suited for the complex anatomy of tumors and sensitive peripheral organs [33]. Although protons have a biological activity comparable to photons, it is higher for heavy ions [34]. In particular, carbon ion therapy has been found to perform better in resistant tumors near organs at risk [35].

Thus, proton therapy has several advantages over photon-based approaches in the treatment various tumor types while minimizing the dose exposure to critical adjacent tissues. Carbon ions are also known to have similar dosimetric advantages as protons. On the other hand, the former ions are biologically more powerful than than the latter. Therefore, there is an increasing interest in carbon ion therapy (CIT) in the treatment of cancer types that are biologically aggressive. In addition, there is controversy about new types of ions beyond protons and carbon ions. Current clinical evidence suggests possible advantages of CIRT over state of the art photon or proton therapy in radiation resistant tumors.

## 5. Conclusion

This study compared proton and carbon beams on the water phantom using Bragg curves and the lateral straggle energy doses. The findings revealed that the proton beam reached the same range with carbon, with an average of 22 times less energy than the proton beam compared to the carbon beam. In other words, the proton beam is more advantageous because it has a greater range with less energy. The input LET is on average 1.554 eV/A in the proton beam, which is 23 times lower than carbon. Bragg transferred an average of 36 times more LETs at peak amplitude than carbon protons. The larger the Bragg peak amplitude of the carbon beam than the proton, the more energy it transfers to the tumor, which means more tumor control. The FWHM value of carbon is about 63% narrower than the proton. In other words, it has a sharper Bragg peak value. The proton's penumbra value is about 53% larger than carbon. In lateral straggle, the proton scattered approximately 1.32 mm, in other words 70% more than the carbon. As the energy increases, the proton beam scatters approximately 1.89 mm, with a 64% increase, while carbon beam and scatters 0.55 mm, with a 59% increase.

It can be argued that the proton beam is more attractive, considering that the management becomes difficult and the operating cost is higher as the energy increases in the accelerators. However, considering the high dose transfer and low lateral straggle results, we can suggest that carbon is more advantageous in the treatment of tumors close to critical tissues.